\documentstyle[epsfig,prl,aps,floats]{revtex}
\tighten

\begin{document}
\draft

\title{Gravitational waves from inspiraling compact binaries: \\
       Second post-Newtonian waveforms as search templates II}
\author{Serge Droz}
\address{Department of Physics, University of Guelph, Guelph,
         Ontario, N1G 2W1, Canada}
\wideabs
{\maketitle
\begin{abstract}
We present further evidence that the second post-Newtonian (pN) 
approximation to the gravitational waves emitted by inspiraling 
compact binaries is sufficient for the detection of these systems. 
This is established by comparing the 2-pN wave forms to signals 
calculated from black hole perturbation theory. Results are presented 
for different detector noise curves. We also discuss the validity of 
this type of analysis.
\end{abstract}
\pacs{Pacs numbers: 04.25.Nx, 04.30.Db, 04.80.Nn}}

\narrowtext

%
%

\section{Introduction}

There is currently a large effort underway to build an international 
network of interferometric gravitational wave detectors. These include 
the two American LIGO detectors, the French-Italian VIRGO detector, 
the somewhat smaller British-German GEO600 and the Japanese TAMA 
detector.

One of the most promising sources for detection are compact 
inspiraling binary systems \cite{Thorne:1987}. These systems, 
consisting of a pair of black holes or neutron stars, lose energy 
through the emission of gravitational radiation and therefore spiral 
inward. The emission is strongest during the last few minutes of the 
inspiral; just before the orbit becomes unstable and the two objects 
plunge towards each other. For stars with masses a few times that of 
the sun this last burst of waves falls within the frequency band of 
the previously mentioned detectors.

The method of choice to search for such signals in a noisy data stream 
is {\em matched filtering} \cite{Wainstein:1962}, where the detector 
output is compared to a bank of templates. A good overlap or match 
then indicates the presence of a signal. It can be shown 
\cite{Wainstein:1962} that matched filtering is an optimal method. 
However, to use this technique an accurate knowledge of the expected 
signals is necessary to set up the template bank. Since no exact 
solutions of Einstein's equations are available which describe a 
binary inspiral, we have to resort to some approximation scheme. The 
templates used by LIGO are calculated using a post Newtonian (pN) 
expansion in the orbital velocity of the binary 
\cite{Blanchet:1996pi,Will:1996zj,Blanchet:1995fg}. However, since the 
pN expansion converges slowly \cite{Poisson:1995vs}, if at all, there 
is a need to ascertain the quality of these templates.

How does one assess the quality of these templates? In a first paper 
\cite{Droz:1997fk} we compared the 2-pN templates to waveforms 
calculated from black hole perturbation theory 
\cite{Poisson:1993zr,Teukolsky:1973}. It was found that the 2-pN 
templates are sufficiently accurate for the detection of a binary 
signal, but not for parameter estimation. However, this first paper 
did not elaborate on the validity of the analysis presented. In this 
publication we address this question, and also present results for a 
wider selection of detector noise curves.

The paper is organized as follows. In section \ref{se:II} we introduce 
the notation and briefly present the formalism used in the remainder 
of the paper. Specifically we introduce the fitting factor 
\cite{Apostolatos:1996rf} which is a measure of the quality of a 
template set. In section \ref{se:valid} we address some of the 
problems encountered when calculating the fitting factor using black 
hole perturbation theory. Finally, in section \ref{se:results} we 
present our results, and finally in section \ref{se:conclude} we 
summarize our conclusions.

%
%

\section{Assessing the quality of a template set}
\label{se:II}

\subsection{The fitting factor}
Let us consider the output of a gravitational wave detector of the form
\begin{equation}
  s(t) = h(t,\vec{\mu}) + n(t).
  \label{eq:outpout}
\end{equation}  
Here $n(t)$ denotes the detector noise and $h(t,\vec{\mu})$ is the 
(possibly absent) signal, which depends on a set of parameters 
$\vec{\mu}$. We define the {\em signal to noise ratio} (SNR) 
\cite{Cutler:1994ys} associated with the template $t_i$ by
\begin{equation}
  \rho_i = \frac{( s | t_i )}{{\rm rms} ( n | t_i )},
  \label{eq:snr1}
\end{equation}
where the inner product $( \cdot | \cdot )$ is defined by
\begin{equation}
  ( s | t ) := 2 \int^\infty_0 \frac{df}{S_n(f)}
     \left(\hat{s}(f) \hat{t}^*(f) +  \hat{s}^*(f) \hat{t}(f) \right).
  \label{eq:innerp}
\end{equation}
As usual hats denote the Fourier transform $\hat{s}(f) = \int e^{-2 
\pi i f t} s(t) dt$ and an asterix indicates complex conjugation. The 
function $S_n(f)$ is the noise's one sided spectral density.  

In this work we use an analytic fit to the noise curves of the various 
detectors \cite{Abramovici:1992ah,Flanagan:1998sx}:
\begin{equation}
  S_n(f) =  S_0 \left( S_1 + 4 \left( \frac{f}{f_{\rm m}} \right)^2 
                          + 2 \left( \frac{f}{f_{\rm m}} \right)^{-4}
						  + 2 \left( \frac{f}{f_1} \right)^{-18} \right) 					 
  \label{eq:So}	
\end{equation}
for $f > f_r$ and $S_n(f) = \infty$ otherwise. Table \ref{tb:S} lists 
the parameters for a few different detectors.
\begin{table}[ht]
\begin{tabular}{lccccc}
 Detector      & $S_0   $   & $S_1 [{\rm Hz}^{-1}]$  & $f_{\rm m}$[Hz]   & $f_1$[Hz] & $f_r$[Hz]  \\ \hline
 Caltech 40m   & $0.5 \times 10^{-42}  $ & $10.0$ & $610.0$ & $183.0$ & $140.0$ \\ 
 initial LIGO  & $0.5 \times 10^{-46}  $ & $4.0$ & $200.0$ & $0.0$ & $40.0$\\ 
 advanced LIGO & $1.5 \times 10^{-49}  $ & $4.0$ & $ 70.0$ & $0.0$& $10.0$ \\ 
\end{tabular}

\caption{\label{tb:S} The parameters used in the analytic fit of the 
various detector noise curves. Note that the value of $S_0$ is not 
important for the calculation of the fitting factor.}
\end{table}

For Gaussian noise \cite{Cutler:1994ys} the expectation value
of the SNR is given by
\begin{equation}
   \langle \rho_i \rangle = \frac{(h | t_i)}{\sqrt{(t_i|t_i)}}.
   \label{eq:exSNR}
\end{equation}
This expression is maximized for $t_i = h$, i.e.~for the Wiener optimal
filter. We can thus rewrite equation (\ref{eq:exSNR}) as 
\begin{equation}
 \langle \rho_i \rangle = \frac{(h | t_i)}{\sqrt{(t_i|t_i) (h|h)}} \, \langle \rho
 \rangle_{\rm max}.
   \label{eq:eSNR}
\end{equation}
The quantity 
\[
      {\cal A}_i = \frac{(h | t_i)}{\sqrt{(t_i|t_i) (h|h)}}
\]
is called the ambiguity function. It gives the loss of SNR due to an 
inaccurate template $t_i$. Its maximum $FF = \max_i {\cal A}_i$ is the 
fitting factor \cite{Apostolatos:1996rf}. It gives the loss of SNR due 
to a non optimal set of templates.

Obviously to calculate the fitting factor we need to know the signal 
$h$ we are looking for. In this paper we use a signal calculated from 
black hole perturbation theory as a substitute for the ``real'' signal 
$h$ \cite{Poisson:1993zr}. Black hole perturbation theory is a priori 
only applicable in the small mass ratio limit. However, we believe 
that black hole perturbation theory can give useful information for 
larger mass ratios. We elaborate on this statement below.

Let us now discuss the signals and templates in some detail.

\subsection{The reference signal}

Black hole perturbation theory \cite{Poisson:1993zr} gives the two 
independent components of the gravitational wave field as
\begin{equation}
  h_+(t) - i h_{\times}(t) \propto \sum_{l=2}^{\infty} \sum_{m=-l}^{l=m}
  A_{lm}(v) \, e^{i m \Omega t} {}_{-2}Y_{lm}(\vartheta,\varphi).
  \label{eq:fixed}
\end{equation}  
Here the functions $A_{lm}(v)$ are the mode amplitudes which are 
calculated numerically by solving the Teukolsky equation. The 
functions ${}_{-2}Y_{lm}(\vartheta,\varphi)$ are spin-weighted 
spherical harmonics \cite{Goldberg:1967}; the angles $\vartheta$ and 
$\varphi$ refer to a Cartesian system centered at the black hole with 
the $z$-axis orthogonal to the orbital plane. The orbital velocity $v$ 
is related to the angular velocity $\Omega$ by the relation $v^3 = M 
\Omega$, where $M=m_1+m_2$ is the total mass.

This waveform describes a particle moving in a fixed circular orbit 
around a Schwarzschild black hole. We can easily generalize this 
expression to an adiabatic inspiral (see Ref. \cite{Droz:1997fk} for 
details). To simplify expression (\ref{eq:fixed}) we choose $\varphi = 
0$ (since the orbit is circular this represents no loss of generality) 
and a detector which singles out the $+$ polarization. Finally, we 
calculate the Fourier transform by means of a stationary phase 
approximation. We obtain
\begin{equation}
  \hat{h}(f) \propto  \sum_{l=2}^{\infty} \sum_{m=2}^{l=m}
    B_{lm}(v) e^{i m \psi(v)} S_{lm}(\vartheta).
   \label{eq:spa}
 \end{equation}
The functions $B_{lm}(v)$ and $\psi(v)$ are calculated numerically 
(see Ref. \cite{Droz:1997fk} for details) and $S_{lm}(\vartheta) = 
{}_{-2}Y_{lm}(\vartheta,0) + (-1)^l {}_{-2}Y_{l-m}(\vartheta,0)$. In 
this expression the gravitational wave frequency is related to the 
orbital velocity by $f = m v^3 / 2 \pi M$. Black hole perturbation 
theory predicts that the innermost stable circular orbit (ISCO) is 
located at $v_{\rm ISCO}=1/\sqrt{6}$, after which $h(v) \equiv 0$. In 
principle, the sum over the index $l$ extends to infinity. We have 
numerically verified that in practice we can truncate the series at 
$l=3$.

\subsection{The templates}
   	
As was mentioned previously, the templates to be used for LIGO data 
analysis are the restricted 2-pN waveforms \cite{Blanchet:1996pi}. 
(The term restricted means that we keep the phase accurate to second 
pN order but use only the leading order amplitude.) Using the 
stationary phase approximation gives
\begin{equation}
  \hat{t_i} := \hat{t}(f, \vec{\theta})  
     \propto  f^{-\frac{7}{6}} e^{i \psi(f)},
  \label{eq:ti}
\end{equation}  
where $\psi(f)  =  2 \pi f t_c - \phi_c + \phi(f)$ and
\begin{eqnarray}
\phi(f) & = & \frac{3}{128} 
  (\pi {\cal M} f)^{-5/3} \left[1 + \frac{20}{9}
\left(  \frac{743}{336} + 
 \frac{11}{4} \eta \right) x^2 - 16\pi x^3
 \right.   \nonumber 
    \\ & & \mbox{} \left.  
  + 10 \left(  \frac{3058673}{1016064} + 
 \frac{5429}{1008} \eta 
+  \frac{617}{144} \eta^2 \right) x^4 \right], 
\label{eq:pnPhase} 
\end{eqnarray}
with $x^3 = \pi \eta^{-3/5} {\cal M} f$. Here, $\vec{\theta} = 
\{\phi_c, t_c, {\cal M}, \eta\}$ are the template parameters, with 
$\phi_c$ the phase at coalescence (formally $f=\infty$), $t_c$ the 
time at coalescence, ${\cal M} = (m_1 m_2)^{\frac{3}{5}} 
M^{-\frac{1}{5}}$ the chirp mass, and $\eta = m_1 m_2 / M^2$ the 
mass-ratio parameter. Note that $\eta$ is maximal for for $m_1 = m_2$ 
with a value of $0.25$.

%
%

\section{Validity of the test mass approximation}
\label{se:valid}
Black hole perturbation theory is valid only for small values of 
$\eta$, say $\eta < \eta_{\rm max} = 0.1$. This poses a problem, since 
most systems we are interested in have a mass ratio of $\eta > 0.2$. 
Furthermore, for a fixed chirp mass, binaries with small mass ratios 
are more relativistic and we expect the pN templates to perform 
poorly.

We can estimate of the minimal value that $\eta$ should take for the 
pN approximation to be valid. We expect pN corrections in the phase 
(\ref{eq:pnPhase}) to be small if $x < x_0 = 0.25$. Then, for fixed 
chirpmass ${\cal M}$ (i.e.~for a fixed number of wave cycles), we find
\[
  \eta_{\rm min} = x^{-5}_0 ( \pi f {\cal M})^\frac{5}{3}
\]
at a given frequency $f$. For example, for ${\cal M}= 1M_\odot$, $f = 
200$ Hz (the peak sensitivity of the initial LIGO detector) and $x_0 = 
0.25$ we get $\eta_{\rm min} = 0.06$. Thus for $\eta \in [0.06, 0.1]$ 
we expect out analysis to give valid results.

\begin{figure}[ht]
\epsfxsize=8cm
\epsffile{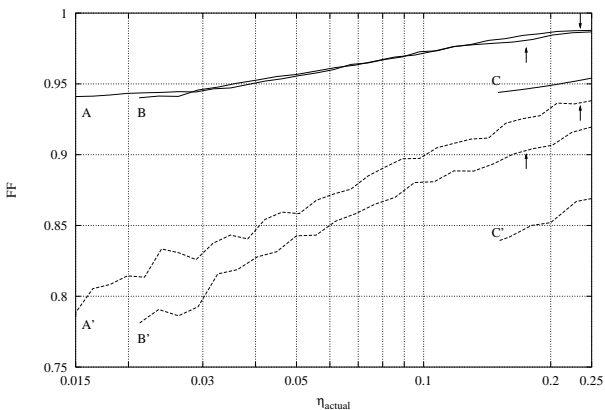}
\caption{\label{fig:ff} shows the fitting factor as a function of 
   $\eta_{\rm actual}$ (i.e.~the signal's $\eta$) for constant values of 
   the chirp mass ${\cal M}$ for the 40 meter noise curve. The three 
   systems shown in this plot correspond to chirp masses ${\cal M}_A = 
   0.87 M_\odot$, ${\cal M}_B = 1.22 M_\odot$ and ${\cal M}_C = 3.92 
   M_\odot$. The primed labels correspond to signals which include all 
   modes $l=2,3, |m|<l$, whereas the unprimed labels correspond to 
   signals only consisting of the (dominant) $l=m=2$ mode. The arrows 
   indicate where the respective curves leave the targeted mass range 
   $0.8 M_\odot < m < 3.2 M_\odot$ of the detector. }
\end{figure}

However we still cannot say anything about systems with mass ratios 
larger than $\eta > 0.1$. Furthermore, for some noise curves 
$\eta_{\rm min} > \eta_{\rm max}$. For example, at the peak 
sensitivity of the 40-meter noise curve ($f = 610$ Hz) we get 
$\eta_{\rm min} > 0.25$.

To estimate how important the corrections to the fitting factor due to 
non-zero $\eta$ are, we can set $\eta=0$ in the coefficients of $x$ in 
the phase (\ref{eq:pnPhase}). This essentially gives us a 2-pN 
approximation to the perturbation theory signal. Somewhat surprisingly 
we find, within numerical accuracy, no difference in the fitting 
factors calculated for the two types of templates. This strongly 
suggests that the non-zero corrections in (\ref{eq:pnPhase}) play no 
essential role, and that our analysis stays valid even for $\eta > 
\eta_{max}$.

Since the binary system becomes less relativistic with increasing 
$\eta$ and constant chirp mass we expect the fitting factor to 
increase as well. This provides us with an additional consistency 
check. This behaviour is indeed observed (see Figure \ref{fig:ff}).

\section{Results}
\label{se:results}

Calculating the fitting factor is in principle straightforward. 
However evaluating the inner product (\ref{eq:innerp}) numerically can 
be tricky. Two methods were used to numerically calculate the fitting 
factor. In the first we calculated the integral in (\ref{eq:innerp}) 
using a Romberg integration \cite{William:1995}. The second method 
uses the fact that $\hat{h}(f) \propto \exp{2 \pi t_{c} f}$ and thus 
the overlap integral (\ref{eq:innerp}) can be evaluated using FFTs. 
Signal and templates in the latter case were calculated using the 
GRASP library \cite{Allen:1997}. For all relevant cases the two 
methods give the same answer within a few percent. As it turns out, 
the second method is a lot faster despite the many more ``$t_{c}$'' 
values that are calculated.

The ambiguity function depends a priory on the four parameters 
$\phi_{c}, t_{c}, {\cal M}$ and $\eta$. Maximization over the phase 
$\phi_{c}$ is trivial. If FFTs are used maximization over $t_{c}$ is 
also s trivial. This leaves us with a reduced ambiguity function to 
maximize, depending on three or two variables, respectively. In the 
first case maximization is difficult, because ${\cal A}(t_{c},{\cal 
M},\eta)$ possesses many local maxima. These seem to mostly disappear 
after maximization over $t_{c}$, so maximizing the ambiguity function 
using FFTs is much easier.

In a first run, reported in \cite{Droz:1997fk}, we calculated fitting 
factors for the restricted 2 pN templates assuming the advanced LIGO 
noise curve. We found that the templates are generally adequate for 
detection, but not for parameter estimation. We now have extended the 
analysis to also cover the Caltech 40 meter prototype and the initial 
LIGO noise curves. Our results are summarized in Table \ref{tb:res} 
for a few selected cases. For each detector there is at least one 
system with a fitting factor larger than $90\%$ which is deemed 
acceptable by Apostolatos \cite{Apostolatos:1996rf}.

\begin{table}[ht]
 \begin{tabular}{ccccc} 
              & \multicolumn{2}{c}{Advanced LIGO} & initial LIGO & 40-meter \\
	Templates &	 Newtonian & 2 pN                 &   2 pN    &  2 pN \\ \hline
    $ 1.4 -  1.4 M_\odot$ & $97.2\%$ & $93.0\%$ & $95.8\%$ & $94.0\%$ \\ 
	$ 0.5 -  5.0 M_\odot$ & $51.6\%$ & $95.2\%$ & $89.7\%$ & $89.0\%$ \\ 
	$ 1.4 - 10.0 M_\odot$ & $55.8\%$ & $91.4\%$ & $88.4\%$ & $< 85.0\%$ \\ 
	$10.0 - 10.0 M_\odot$ & $70.1\%$ & $91.7\%$ & $86.3\%$ &  \\ 
	$ 4.0 - 30.0 M_\odot$ & $61.3\%$ & $86.6\%$ & $67.9\%$ &  \\ 
  \end{tabular}
  \caption{\label{tb:res} The fitting factors $FF$ for selected binary 
  systems and for various noise curves.}
\end{table}  

\begin{figure}[ht]
\epsfxsize=8cm
\epsffile{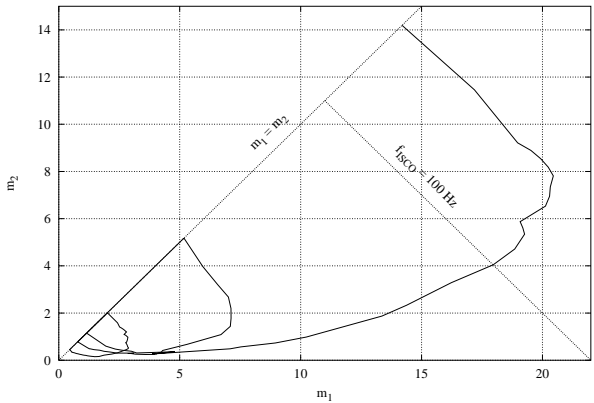}
\caption{\label{fig:cont} shows the contours of $FF=90\%$ in the $m_1 
- m_2$ plane for the Caltech 40 meter prototype, the initial and 
advanced LIGO detectors. Units are in solar masses. The dotted curve 
to the right denotes the line where the frequency at the inner most 
stable orbit (ISCO), as predicted by perturbation theory, becomes less 
than $100 $Hz. While there are still templates with $FF>90\%$ in this 
region, the total SNR for these systems is very low since they spend 
very little time in the advanced LIGO frequency band. }
\end{figure}

A more complete picture is given by Figure \ref{fig:cont} which shows 
the contours of $FF = 90\%$ in the $m_1 - m_2$ plane for the 40 meter, 
initial and advanced LIGO detectors. Since the frequency band for the 
advanced LIGO starts at $10 $Hz we have to follow a large number of 
wave cycles, especially for low mass systems. This makes the 
evaluation of the fitting factor difficult in terms of available 
computer power.

In a next step we calculated the fitting factor as function of $\eta$ 
for fixed chirp mass ${\cal M}$. (These are the mass parameters for 
the signal and they should not be confused with the mass parameters in 
the ambiguity function ${\cal A}$.) As expected, the fitting factor 
drops considerably as $\eta$ decreases. Figure \ref{fig:ff} shows the 
curves $FF(\eta)$ for three different values of the chirp mass for the 
40 meter noise curve. Note that the end points of these curves lie 
well outside the mass regime considered of interest for this detector. 
Similar results were obtained for the advanced and initial LIGO 
detectors.
\begin{figure}[ht]
\epsfxsize=8cm
\epsffile{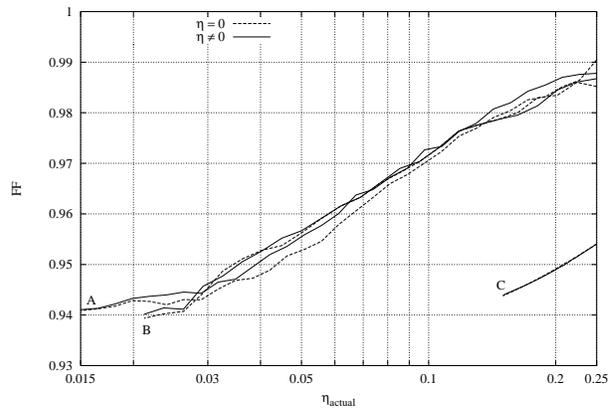}
\caption{\label{fig:ff2} shows plots of the fitting factor $FF$ as a 
function of $\eta$ with constant chirp mass ${\cal M}$ for the same 
three systems. The curves shown were calculated using test mass 
signals consisting of the $l=m=2$ mode only. The dashed curves 
correspond to templates with $\eta=0$. The solid lines represent 
proper 2 pN templates. There is practically no difference between the 
two cases, indicating that perturbation theory gives valid results 
even for relatively large mass ratios.}
\end{figure}

Finally, we repeated this calculation with $\eta=0$ in the 
coefficients of the phase (\ref{eq:pnPhase}). The values for the 
fitting factor were found to agree with the previous values within 
numerical accuracy. Figure \ref{fig:ff2} compares the two cases for 
the 40m noise curve for the $l=m=2$ signal. We take this as an 
indication that perturbation theory waveforms are adequate to test the 
performance of a set of templates, even when $\eta$ is not small.

%
%

\section{Conclusions}
\label{se:conclude}
The results of this analysis suggest that the 2 pN templates are 
adequate for the detection of gravitational radiation emitted from 
inspiraling compact binary systems, when the masses lie in the region 
shown in Figure \ref{fig:cont}. We emphasize that we do not claim that 
the 2 pN templates are all we need to know for a LIGO-type search for 
compact binaries. Clearly, more accurate templates are needed to 
detect the larger mass systems as well as for parameter estimation 
\cite{Droz:1997fk}. But we do believe that the 2-pN templates give us 
a good chance to detect gravitational wave from, at least, some 
systems.

While it is not a priory clear that black hole perturbation theory 
makes a good tool to ascertain the quality of 2-pN templates for large 
mass ratio systems, we have given arguments to support our results. At 
the very least, we feel our analysis gives a qualitatively correct 
picture.

%
%
\section*{Acknowledgments}

The authour would like to thank Eric Poisson for his stimulating 
comments. This work was supported by the Natural Sciences and 
Engineering Research Council of Canada.

%
%


\end{document}